\documentclass[prb,preprint]{revtex4}
\usepackage{amsmath}

\usepackage{amssymb}
\usepackage{graphicx}

\begin{document}

\title{Numerical approach to the Schr\"{o}dinger equation in momentum space}
\author{William A. Karr}
\affiliation{Department of Physics, 
Indiana University Purdue University Indianapolis (IUPUI), 
Indianapolis, Indiana 46202, USA}
\author{Christopher R. Jamell}
\affiliation{Department of Physics, 
Indiana University Purdue University Indianapolis (IUPUI), 
Indianapolis, Indiana 46202, USA}
\author{Yogesh N. Joglekar}
\email{yojoglek@iupui.edu}
\affiliation{Department of Physics, 
Indiana University Purdue University Indianapolis (IUPUI), 
Indianapolis, Indiana 46202, USA}


\begin{abstract}
The treatment of the time-independent Schr\"{o}dinger equation in real space is 
an indispensable part of introductory quantum mechanics. In contrast, the 
Schr\"{o}dinger equation in momentum space is an integral equation that is not 
readily amenable to an analytical solution, and is rarely taught. We present a 
numerical approach to the Schr\"{o}dinger equation in momentum space. After a 
suitable discretization process, we obtain the Hamiltonian matrix and diagonalize 
it numerically. By considering a few examples, we show that this approach is 
ideal for exploring bound states in a localized potential, and complements the 
traditional (analytical or numerical) treatment of the Schr\"{o}dinger equation 
in real space. 
\end{abstract}

\maketitle

\section{Introduction}
\label{sec:intro}
The treatment of the time-independent Schr\"{o}dinger equation for a 
non-relativistic particle of mass $m$ is a prime element of quantum mechanics 
courses.\cite{introqm,merz} The treatment of this second-order differential 
equation introduces students to the effect of boundary conditions on 
quantization\cite{introqm} and to the Sturm-Liouville problem.\cite{sl1,sl2} For 
only a few potentials $V(x)$ can the Schr\"{o}dinger equation 
\begin{equation}
\label{eq:se}
-\frac{\hbar^2}{2m}\frac{d^2}{dx^2}\psi_\alpha(x)+ V(x)\psi_\alpha(x)=
E_\alpha\psi_\alpha(x)
\end{equation}
be solved analytically, and the eigenvalue spectrum $E_{\alpha}$ and the 
complete set of orthonormal eigenfunctions $\psi_\alpha(x)$ be obtained. 
Introductory texts typically include a quantum well or step, a harmonic 
oscillator, a delta function potential, and various combinations.\cite{introqm} 
The number of analytically solvable potentials is even smaller in higher 
dimensions. Notable exceptions are those with a central potential where 
rotational invariance allows us to obtain a second-order differential equation 
for the radial wavefunction in an effective potential that takes into account the 
centripetal barrier.\cite{introqm,merz} Although such an equation is not, in 
general, analytically solvable, it is a significant improvement over the 
second-order partial differential equation.

To explore the bound states in an arbitrary potential $V(x)$, we can use 
the Wentzel-Kramers-Brillouin (WKB) approximation\cite{merz} which provides a 
semiclassical picture of quantized energy eigenvalues. Another approach 
is to discretize Eq.~(\ref{eq:se}) and obtain the matrix 
equation
\begin{equation}
\label{eq:discretese}
\sum_{j=-N}^{N}\left[-\frac{\hbar^2}{2m}D_{ij}+V(x_i)\delta_{ij}\right]
\psi_\alpha(x_j)=\sum_{j=-N}^{N}H_{ij}\psi_\alpha(x_j)=E_\alpha\psi_\alpha(x_i)
\end{equation} 
where $x_j= j\Delta x$, $\Delta x$ is the spacing between adjacent points 
along the discretized $x$-axis, and $N\Delta x=X_c$ denotes the spatial 
cutoff chosen 
such that $X_c\gg a$ where $a$ is the characteristic length scale of the 
potential $V(x)$. The tridiagonal second-derivative matrix has entries 
$D_{ij}=\left[\delta_{i,j-1}-2\delta_{ij}+\delta_{i,j+1}\right]
/(\Delta x)^2$. In principle, as $X_c\rightarrow\infty$ and 
$\Delta x\rightarrow 0$, the eigenvalues and eigenvectors of the $(2N+1)
\times (2N+1)$ matrix $H_{ij}$ approach the spectrum of the original 
continuum problem. However, due to the diverging prefactor $(\Delta x)^{-2}$ 
and the error in the matrix $D_{ij}$ at the end-points $\pm X_c$, the matrix 
diagonalization approach does not lead to stable continuum results. Instead, 
the Numerov method has to be used to numerically obtain the eigenvalues 
and eigenfunctions.\cite{numerov,landau} Another approach is to use the 
eigenfunction expansion method,\cite{ee} which results in a matrix equation 
for the expansion coefficients.\cite{eemodern} 

In this paper we show that the stability and convergence 
issues\cite{numerov} are circumvented by the Schr\"{o}dinger equation in 
momentum space. In Sec.~II we review the equation and the 
corresponding discretized-matrix eigenvalue problem. This method is ideal 
for localized potentials with a finite Fourier transform. We 
present the bound-state spectra for a few well-known potentials and compare 
them with analytical results whenever possible. In Sec.~\ref{sec:hd} we 
discuss the generalization of our approach to the Schr\"{o}dinger equation in 
higher dimensions.\cite{landau} We conclude in 
Sec.~\ref{sec:disc} with a discussion and suggested problems. The method 
presented here is complementary to the standard differential-equation 
approach, can be explored in introductory quantum mechanics courses, and is 
accessible to junior or senior undergraduate students familiar with Matlab, 
Maple, Mathematica, or LAPACK.

\section{Schr\"{o}dinger equation in momentum space}
\label{sec:pspace}
We start with the Fourier transform of the one-dimensional Sch\"{o}dinger 
equation, Eq.~(\ref{eq:se}),
\begin{equation}
\label{eq:pspace}
\epsilon_{p}\psi_\alpha(p)+\!\int_{-\infty}^{\infty}\frac{dp'}{(2\pi\hbar)} 
V(p-p')\psi_\alpha(p')=E_\alpha\psi_\alpha(p).
\end{equation}
Here $\psi_\alpha(p)$ is the momentum-space wavefunction, which represents the 
probability amplitude for the particle to have momentum $p$, $\epsilon_{p}=p^2/2m$ 
is the (non-relativistic) kinetic energy of the particle, and $V(q)$ is the 
Fourier transform of the external potential $V(x)$. We use the same symbol for 
the eigenvector $|\psi_\alpha\rangle$ and the potential energy operator $\hat{V}$ 
in both real and momentum space; the fact that $\psi_\alpha(x)$ and 
$\psi_\alpha(p)$ are different functions is understood.\cite{dirac,sakurai} The 
integral equation \eqref{eq:pspace} has been used to study the scattering 
problem\cite{sakurai,scatter} and the bound state in a $\delta$-function 
potential.\cite{onedelta} 

To convert Eq.~(\ref{eq:pspace}) into a form suitable for numerical exploration, 
we use $a_0$ to denote the length-scale, define the momentum scale by 
$\hbar/a_0$, and use $E_0=\hbar^2/2ma_0^2$ as the unit of energy. Because 
$a_0$ is arbitrary in the continuum limit, the spectrum with a potential 
characterized by depth $V_0$ and range $a$, $V(x)=V_0f(x/a)$, is determined by 
the dimensionless parameter $V_0/E_a= V_0/(\hbar^2/2ma^2)$. For the numerical 
calculations it will be useful to choose $V_0/E_0=1$ and vary $a/a_0$ to access 
various values of $V_0/E_a$. In terms of the dimensionless variables 
Eq.~(\ref{eq:pspace}) leads to a matrix equation after discretization:
\begin{equation}
\label{eq:discretepspace}
\left[u^2_n\delta_{mn}+\tilde{V}_{mn}\right]\psi_\alpha(u_n)=
H_{mn}\psi_\alpha(u_n)=\tilde{E}_\alpha\psi_\alpha(u_m),
\end{equation}
where a sum over the repeated index $n$ is understood. Here 
$u_n=n\Delta u=n\Delta p a_0/\hbar=p_n a_0/\hbar$ is the dimensionless 
momentum, $\Delta u$ is the spacing between adjacent points along the 
$u$-axis, $\tilde{E}_\alpha=E_\alpha/E_0$ is the dimensionless eigenvalue, and 
$\tilde{V}_{mn}= V(p_m-p_n)\Delta u/(2\pi E_0 a_0)$ is the dimensionless 
potential along with the discrete measure $\Delta u/(2\pi)$. Thus, the 
integral Schr\"{o}dinger equation (\ref{eq:pspace}) has been recast as a matrix 
eigenvalue problem, Eq.~(\ref{eq:discretepspace}). The eigenvalues and 
eigenvectors of the dimensionless Hamiltonian matrix $H_{mn}$ are 
obtained using standard software packages. The results presented here were 
obtained by using Matlab and were verified by using LAPACK. This discretization 
process involves some computational subtleties that we discuss in the following, 
but provides an excellent way to study bound states in potentials that are 
localized in real space. 

The size of the Hamiltonian matrix $H_{mn}$ is determined by the dimensionless 
ultraviolet momentum cutoff $U_c=P_c a_0/\hbar$, where $P_c$ is upper limit of 
the integration range in Eq.~(\ref{eq:pspace}), and the dimensionless spacing 
$\Delta u$. Note that $2\pi a_0/\Delta u$ and $2\pi a_0/U_c$ impose the upper 
and lower limit respectively on the real-space size of the bound-state 
wavefunction. Thus, they need to be chosen for a given potential so as to obtain 
results that are valid in the continuum limit, $U_c\rightarrow\infty$ and 
$\Delta u\rightarrow 0$. If the external potential $V(x)$ is even, the 
Hamiltonian $H_{mn}$ is real. Therefore, the eigenfunctions $\psi_\alpha(p)$ 
have a definite parity and are real.\cite{merz} In this case it is 
sufficient to restrict ourselves to positive momenta $m,n\geq 0$.

To demonstrate these considerations, we start with a quantum well with depth 
$V_0$ and width $a$ centered at the origin, $V(x)=-V_0\theta(a-2|x|)=V(-x)$ 
where $\theta(x)$ is the Heavy-side function.\cite{sl1,sl2} In this case the 
real Hamiltonian matrix is given by
\begin{equation}
\label{eq:pbox}
H_{mn}=u^2_n\delta_{mn}-\frac{\Delta u}{2\pi}\frac{2aV_0}{a_0E_0}\frac{
\sin\left[(u_n-u_m)a/a_0\right]}{\left[(u_n-u_m)a/a_0\right]}=H_{nm}.
\end{equation}
The factor of $2$ in the potential matrix elements arises from the 
restriction $m,n\geq 0$. The one-dimensional $\delta$-function potential, 
$V(x)=-\lambda_1\delta(x)$, is obtained as a 
limiting case when $a\rightarrow 0$, $V_0\rightarrow\infty$ with 
$aV_0=\lambda_1$. In this limit for an attractive potential, $\lambda_1>0$, 
the system has one exponentially bound state 
$\psi_b(x)=\sqrt{\kappa}\exp(-\kappa |x|)$ with size 
$\kappa^{-1}=\hbar^2/m\lambda_1$ and energy\cite{introqm,merz} 
$E_b=-m\lambda_1^2/2\hbar^2$.

We obtain the eigenvalues and eigenvectors of the matrix $H_{mn}$ for 
$0\leq|\lambda_1/E_0a_0|\leq 1$ with two different values of 
$\Delta u=$\{0.01,0.005\} and two different cutoffs $U_c=$\{10,20\}. The 
corresponding matrix dimension $N=U_c/\Delta u$ in these four cases varies from 
$1000\times 1000$ to $4000\times 4000$. We find that the energy spectrum has one 
negative eigenvalue and the positive eigenvalues form a quadratic band 
representative of a free particle. (The positive-energy unbounded states are not 
accessible via the Numerov method.\cite{numerov}) 
Figure~\ref{fig:1delta}(a) shows that the bound-state energy $E_b(\lambda_1)$ 
matches the analytical result. Figure~\ref{fig:1delta}(b) shows 
a typical momentum-space wavefunction for the bound-state $\psi_b(p)$ and a 
state with positive energy. As expected, we see that the bound-state 
wavefunction, $\psi_b(p)=2(\hbar\kappa)^{3/2}/(p^2+\hbar^2\kappa^2)$, is broad, 
whereas the positive-energy wavefunction is sharply peaked near a single 
momentum value. We check that the bound-state results are independent of 
$U_c$ and $\Delta u$. (The smallest momentum cutoff is 
chosen such that the contribution from momenta $p>P_c=U_c\hbar/a_0$ to the 
bound-state wavefunction is negligible.) Increasing $U_c$ affects 
the eigenvalues and eigenvectors near the highest energy $E_c/E_0=U_c^2$, 
whereas reducing $\Delta u$ sharpens the momentum-space eigenfunctions at 
positive energies. Thus, this numerical approach is particularly suited to 
study bound states, and may not handle the unbounded positive-energy states 
equally well.

Next, we consider the problem of a deep quantum well $V_0/E_a\gg 1$. We 
diagonalize the matrix $H_{mn}$, Eq.~(\ref{eq:pbox}), with $V_0/E_0=1$, 
$U_c=300$, $\Delta u=0.1$, and $a/a_0=\{20,25\}$. Figure~\ref{fig:qwell} 
shows that the numerically obtained spectrum of the bound-state energies 
measured from the bottom of the quantum well is quadratic, 
$E_n=n^2\pi^2{\cal E}$. This dependence is expected because for an infinite 
quantum well of size $a$ the eigenvalues are given by $E_n=n^2\pi^2 E_a$. The 
prefactor ${\cal E}(U_c,\Delta_u,V_0)\neq E_a$, and a systematic exploration with 
increasing $U_c$ and $a$, and decreasing $\Delta u$ shows that this discrepancy 
is due only to the discretization. We next consider an attractive Gaussian 
potential $V_1(x)=-V_0\exp(-x^2/2a^2)$. In this case a closed-form solution for 
the eigenvalues and eigenfunctions is unknown. The dimensionless Hamiltonian 
becomes 
\begin{equation}
\label{eq:hv2}
H^{(1)}_{mn} = u_n^2\delta_{mn}-\frac{\Delta u}{\sqrt{2\pi}}
\frac{2V_0 a}{E_0 a_0}\exp\left[-\frac{(u_m-u_n)^2a^2}{2a_0^2}\right]=
H^{(1)}_{nm}.
\end{equation}

Figure~\ref{fig:hv2} shows the $a$-dependence of the magnitude of the 
ground-state energy $E_b<0$ obtained by using $V_0/E_0=1$, $\Delta u=0.01$, and 
$U_c=30$. The inset shows the ground state momentum-space wavefunction 
$\psi_G(p)$ for $a/a_0=\{0.1,0.5\}$. As $a/a_0$ increases, the effective 
value of $V_0/E_a$ increases. Thus, the ground state becomes more 
localized in real-space, and the spread of the wavefunction in 
momentum-space increases. 

We emphasize that the bound-state eigenvalues and eigenfunctions obtained from 
the diagonalization of the discrete matrix $H_{mn}$ should be essentially 
independent of the cutoff $U_c\gg 1$ and the spacing $\Delta u\ll 1$, to verify 
that they are valid in the continuum limit 
$U_c\rightarrow\infty$, $\Delta u\rightarrow 0$. Note that even in the limit 
$\Delta u\rightarrow 0$, a finite momentum-cutoff $P_c$ leads to a real-space 
potential that is not the same as the original one:
\begin{equation}
\label{eq:finitevq}
V_c(x)=\!\int_{-P_c}^{P_c}\!\frac{dp}{2\pi\hbar}V(p)e^{ipx/\hbar}\neq V(x).
\end{equation}
Thus the discretization parameters need to be so chosen that the difference 
between $V_c(x)$ and $V(x)$ is negligible. Two typical indicators that the 
continuum limit has not been reached are that some eigenvalues are lower than 
the depth of the potential well, $E_\alpha<-V_0$, and the ground-state 
momentum-space wavefunction is linear, instead of quadratic, near $p=0$. In one 
dimension we can choose the ground state wavefunction $\psi_G(x)$ to be 
positive,\cite{introqm,merz} so that, for an even potential the momentum-space 
wavefunction is 
parabolic at the origin, $\psi_G(p)-\psi_G(p=0)\propto -p^2$. Therefore, a 
linearly varying $\psi_G(p)$ is a clear indication that the bound-state 
eigenvalues and eigenfunctions do not represent continuum results. In the 
following we show that the verification of the continuum limit is 
more subtle in two dimensions and requires a careful treatment.

\section{Numerical approach in higher dimensions}
\label{sec:hd}
For a particle in two or more dimensions, a naive discretization of the integral 
Schr\"{o}dinger equation in Cartesian co-ordinate implies that the Hamiltonian 
matrix has a size $\sim N^D\times N^D$ where $N=U_c/\Delta u$ is the number of 
discrete points along a single axis and $D$ is the dimension. Thus, even in 
two dimensions, the parameters used in Sec.~\ref{sec:pspace} result in 
$10^6\times 10^6$ or larger matrices that are impossible to treat numerically. 
For a central potential in two dimensions, the rotational 
invariance of the Hamiltonian implies that the angular momentum is a good 
quantum number and the eigenfunctions can be labeled by an integer 
angular momentum label $\ell$, $\psi_\alpha({\bf p})=\psi_{\alpha \ell}(p)
\exp(i\ell\theta_p)$, where ${\bf p}=(p,\theta_p)$ is the two-dimensional 
momentum.\cite{introqm,merz,sakurai} The Schr\"{o}dinger equation for a given 
value of $\ell$ becomes\cite{landau} 
\begin{equation}
\label{eq:pspaceD}
\frac{p^2}{2m}\psi_{\alpha l}(p)+\!\int\frac{p'dp'd\theta_{p'}}{(2\pi\hbar)^2}
V(p,p';\theta_p-\theta_{p'})
e^{-i\ell(\theta_p-\theta_{p'})}\psi_{\alpha \ell}(p')=E_{\alpha \ell}\psi_{\alpha \ell}(p)
\end{equation}
where $V(p,p';\theta_{p}-\theta_{p'})=V(|{\bf p}-{\bf p'}|)$ is the 
momentum-space potential and depends only on the angle between 
${\bf p}$ and ${\bf p'}$ due to the central nature of the potential. The 
corresponding dimensionless Hamiltonian matrix becomes 
$H_{mn}(\ell)=u_n^2\delta_{mn}+\tilde{V}_{mn}(\ell)$, where the angular-averaged 
potential matrix is given by 
\begin{equation}
\label{eq:discreteD}
\tilde{V}_{mn}(\ell)=\frac{u_n\Delta u}{2\pi E_0a_0^2} \int^{2\pi}_{0}
\frac{d\theta}
{2\pi}V(u_m,u_n;-\theta)e^{i\ell\theta}.
\end{equation}
Here, $0\leq u_n\leq U_c$ denotes the magnitude of the dimensionless momentum 
and the matrix $H_{mn}(\ell)$ has size $\sim N\times N$. Due to the prefactor 
$u_n\Delta u$ from the two-dimensional area-element in polar co-ordinates, the 
Hamiltonian obeys $H_{nm}(\ell)=(u_m/u_n)H_{mn}^{*}(\ell)$. Thus, the 
discretized Hamiltonian matrix is not Hermitian with respect to transpose of the 
matrix plus complex conjugation. It is Hermitian with respect to the inner 
product defined via the two-dimensional measure. We will focus on $\ell=0$ case 
because for a time-reversal invariant Hamiltonian, the ground state has zero 
angular momentum.\cite{merz,sakurai}

As an illustration, we consider an attractive $\delta$-function potential 
in two dimensions, $V({\bf r})=-\lambda_2\delta^{2}({\bf r})$. Although a 
trivial extension of the one-dimensional problem, it is rarely 
discussed\cite{landau2} in introductory courses, perhaps because the 
bound-state real-space wavefunction is logarithmically 
divergent\cite{twodelta1} in 
the vicinity of the $\delta$-function. The bound-state energy 
$E_b(\lambda_2,U_c)$ depends on the ultraviolet cutoff $U_c$ and has a 
non-analytic dependence on the strength of the potential, $E_b/E_0=-U_c^2
\exp(-4\pi E_0 a_0^2/\lambda_2)$.\cite{landau2,twodelta2} The 
momentum-space Schr\"{o}dinger equation in this case is analytically 
tractable and provides a good test.\cite{twodelta2} Because the 
Fourier transform of this potential is a constant, the Hamiltonian matrix 
becomes
\begin{equation}
\label{eq:2ddelta}
H_{mn}(\ell)=u_n^2\delta_{mn}-\frac{u_n\Delta u}{2\pi}\frac{\lambda_2}{E_0a_0^2}
\delta_{\ell 0}=\frac{u_n}{u_m}H_{nm}(\ell).
\end{equation}
The $\delta$-function potential affects only $\ell =0$ sector of the Hilbert 
space because wavefunctions with $\ell\neq 0$ vanish at the position of the 
$\delta$-function due to the centripetal barrier. We verify that there is a 
single bound state for an attractive potential ($\lambda_2> 0$) and none for a 
repulsive potential ($\lambda_2< 0$). Figure~\ref{fig:2delta} shows the 
magnitude of the bound-state energy $E_b(\lambda_2)$ as a function of 
$4\pi E_0 a_0^2/\lambda_2$ for $2\leq\lambda_2/E_0a_0^2\leq 20$. We use 
$\Delta u=0.01$ and two ultraviolet 
cutoffs $U_c=\{10,20\}$. At large values of $\lambda_2/E_0a_0^2\sim U_c$, the 
numerical results deviate from the expected straight-line behavior due to 
discretization. This deviation is systematically suppressed by reducing 
$\Delta u$. Note that for an attractive $\delta$-potential in both one and 
two dimensions the bound-state wavefunction has the same functional form, 
$\psi_b(p)\propto 1/(p^2+\hbar^2\kappa^2)$. However, because the bound-state 
energy $E_{b2}\propto\exp(-1/\lambda_2)$ in two 
dimensions\cite{landau2}, in contrast to the bound-state energy 
$E_{b1}\propto\lambda_1^2$ in one dimension,\cite{introqm,merz} the size 
of the wavefunction in momentum-space in two dimensions is much smaller 
than that in one dimension, $\hbar\kappa_2=\sqrt{2m|E_{b2}}\ll 
\hbar\kappa_1=\sqrt{2m|E_{b1}|}$. We emphasize that the 
dependence of $E_b$ on the cutoff $U_c$ is a peculiar property of the 
weakly bound state in the two-dimensional $\delta$-function potential and 
arises due to the absence of an energy scale in a problem characterized by 
$(\hbar,m,\lambda_2)$. For a general potential, including the attractive 
Coulomb interaction $V(r)=-e^2/r$, we numerically obtain multiple bound-states 
with energies that are independent of the cutoff.\cite{coulomb} 

For a central potential it is straightforward to extend this method to 
higher dimensions. In $D$-dimensions we represent a vector using hyperspherical 
co-ordinates, ${\bf p}=(p,\phi,\theta_1,\theta_2,\ldots,\theta_{D-2})$ where 
$\phi\in [0,2\pi]$ and $\theta_i\in [0,\pi]$.\cite{Ddim} The Hamiltonian is 
then block-diagonalized into blocks with different angular momenta. The 
effective potential in the block $(\ell, \ell_z)$ is obtained by performing an 
integral over angular variables, similar to that in Eq.~(\ref{eq:discreteD}). 
It is not always possible to analytically carry out this integration. The 
resulting Hamiltonian matrix satisfies 
$H_{nm}(\ell, \ell_z)= (u_m/u_n)^{D-1}H^{*}_{mn}(\ell, \ell_z)$, and the resulting 
eigenvectors $\psi_\alpha(u_k)$ are orthogonal with respect to the 
$D$-dimensional inner product
\begin{equation}
\langle\psi_\alpha|\psi_\beta\rangle=\frac{\Delta u}{2\pi}\sum_{k=0}^{N}
\psi^{*}_\alpha(u_k)\psi_\beta(u_k)u_k^{D-1}=0 \qquad (\alpha\neq\beta).
\end{equation}

The exploration of the Hamiltonian matrix $H_{mn}(\ell, \ell_z)$ in $D\geq 2$ 
dimensions emphasizes two important points. First, by explicit 
construction, it generates a set of matrices, each element of which 
appears non-Hermitian and still has a purely real eigenvalue spectrum. 
Second, it explicitly demonstrates that the notion of 
orthonormality and Hermiticity are intimately connected to the inner-product 
used to construct the Hilbert space of wavefunctions.\cite{introqm}

\section{Conclusions}
\label{sec:disc}
We have presented an approach to the real-space 
Schr\"{o}dinger equation via the Hamiltonian matrix in momentum-space that is 
obtained after a suitable discretization.\cite{landau} This method does not 
suffer from the instability associated with discretization of the real-space 
Schr\"{o}dinger equation,\cite{numerov} primarily because the kinetic energy 
term is diagonal in momentum space and, for most physical potentials, the 
amplitude $\tilde{V}_{pp'}$ for scattering from $p$ to $p'$ decays for large 
$|p-p'|$. Therefore, the elements of the matrix $H_{mn}$ near the top-right 
and bottom-left corners are small. 

Our method is best suited for numerically investigating the energies and 
wavefunctions of bound states that occur in a localized central potential 
$V(r)$ with a finite Fourier transform $V(q)$. Many 
well-known examples with confining potentials where all eigenstates are 
localized (an infinite quantum well or a simple harmonic oscillator) cannot 
be studied using our approach because the Fourier transform is ill-defined. 
However, as we have discussed in Sec.~\ref{sec:pspace}, it is possible to 
explore the low-lying eigenstates of such a system by choosing parameters such 
that $V_0/E_a=V_0/(\hbar^2/2ma^2)\gg 1$. Such a deep well, as far as the 
low-lying eigenstates are concerned, can be treated as an infinite well. 

\section{Suggested Problems}

{\it Problem~1}. Obtain the bound-state spectra for the potential 
\begin{equation}
V_\eta(x)=
\begin{cases}
-V_0[1-(2|x|/a)^\eta] & |x|<a/2\\
=0 & \mbox{otherwise},
\end{cases}
\end{equation} 
where $\eta>0$. Note that $V_\eta(x)$ represents a family of potentials 
that extrapolate from a linear ($\eta=1$), a quadratic ($\eta=2$), to a 
quantum well ($\eta\rightarrow\infty$). Choose 
$V_0/E_a=V_0/(\hbar^2/2ma^2)\gg 1$. Compare your results to the WKB 
approximation prediction $E_k=-A_\eta[2k+1]^{2\eta/(2+\eta)}$, where $A_\eta$ 
is a constant and the eigenenergies $E_k$ are measured from the bottom of the 
potential well. 

\medskip {\it Problem~2}. Obtain the analog of Eq.~(\ref{eq:2ddelta}) in three 
dimensions and study the spectrum for a given cutoff $U_c$. Show that a bound 
state arises only when $\lambda_3\geq \lambda_{3c}\sim 1/U_c$, and determine 
$\lambda_{3c}$. Contrast your results with those for a quantum well with 
depth $V_0$ and size $\hbar/P_c=a_0/U_c$.

\begin{acknowledgments}
W.A.K.\ was supported by the IUPUI Undergraduate Research Opportunity Program 
(UROP) through a Summer Fellowship. C.R.J.\ was supported by the National 
Science Foundation GAANN Fellowship. We are thankful to an anonymous referee 
for pointing out Ref.~\onlinecite{landau} to us. 
\end{acknowledgments}

\section*{Figure captions}

\begin{figure}[h!]
\begin{center}
\begin{minipage}{20cm}
\begin{minipage}{9cm}
\hspace{-3cm}
\end{minipage}
\begin{minipage}{9cm}
\hspace{-5cm}
\end{minipage}
\end{minipage}
\caption{\label{fig:1delta}
(color online) (a) Dependence of the magnitude of the bound-state 
energy $|E_b|$ obtained from the matrix $H_{mn}$, Eq.~(\ref{eq:pbox}), on the 
strength $\lambda_1$ of the one-dimensional attractive $\delta$-function 
potential. The energy is in units of $E_0=\hbar^2/2ma_0^2$ and 
$\lambda_1$ is in units of $E_0a_0$. The numerical result (crosses) 
is in excellent agreement with the analytical result\cite{introqm,merz} 
$|E_b|/E_0=\lambda_1^2/4(E_0a_0)^2$ (dashed line). (b) Typical 
momentum-space wavefunctions for the bound state (top curve) and a positive-energy 
state (bottom curve) for $\lambda_1/E_0a_0=0.5$. The momentum $p$ is in 
units of $a_0/\hbar$ and the wavefunction is in units of 
$\sqrt{a_0}$. The width of the bound-state wavefunction $\psi_b(p)$ is given 
by $\hbar\kappa=m\lambda_1/\hbar$. The positive-energy wavefunction is, as 
expected, sharply localized in momentum space.}
\end{center}
\end{figure}

\begin{figure}[h!]
\begin{center}
\caption{\label{fig:qwell}
(color online) Eigenenergies $E_n$ of bound states in a deep quantum well 
$V_0/E_a\gg 1$ obtained from Eq.~(\ref{eq:pbox}). The bound-state energies 
$E_n$ are positive because they are measured from the bottom of the well. 
The energies are in units of 
$E_a=\hbar^2/2ma^2=E_0(a_0/a)^2$ instead of the customary unit $E_0$. The 
solid and the dashed lines represent results 
for $V_0/E_a=(a/a_0)^2=$ 400 and 625 respectively. The dotted line shows the 
analytical result for an infinite quantum well of width $a$, 
$E_n=(n\pi)^2E_a$.}
\end{center}
\end{figure}

\begin{figure}[h!]
\begin{center}
\caption{\label{fig:hv2}
(color online) Magnitude of the ground-state energy $|E_b|$ for a Gaussian 
potential $V(x)=-V_0\exp(-x^2/2a^2)$ as a function of 
$a$ for a fixed depth $V_0/E_0=1$. The energies are in units of 
$E_0$, the length is in units of $a_0$, and the momentum in the 
inset is in units of $\hbar/a_0$. The inset shows the 
ground-state momentum-space wavefunction in units of $\sqrt{a_0}$. 
As $a$ increases, the ground-state wavefunction becomes increasingly 
localized in real space, and is reflected in the broadening of the 
momentum-space wavefunction.} 
\end{center}
\end{figure}

\begin{figure}[h!]
\begin{center}
\caption{\label{fig:2delta}
(color online) Dependence of 
$|E_b|$ on the strength $\lambda_2$ of the attractive two-dimensional 
$\delta$-function potential for ultraviolet momentum cutoffs $U_c=10$ 
(squares) and $U_c=20$ (circles). The energy is in units of $E_0$ 
and $\lambda_2$ is in units of $E_0a_0^2$; $\Delta u=0.01$ 
and $2\leq \lambda_2/E_0a_0^2\leq 20$. The solid lines represent the 
analytical result, $\ln(|E_b|/E_0)=-4\pi E_0a_0^2/\lambda_2 + \ln(U_c^2)$ 
for $U_c=10$ (bottom curve) and $U_c=20$ (top curve). The corresponding 
$y$-intercepts are $\ln(10^2)\approx 4.6$ and $\ln(20^2)\approx 6$. The 
discrepancy between the numerical and analytical results for large 
$\lambda_2/E_0a_0^2\sim U_c$ is decreased when $\Delta u$ is reduced.} 
\end{center}
\end{figure}


\begin{thebibliography}{99}
\bibitem{introqm} D. J. Griffiths, {\it Introduction to Quantum Mechanics} 
(Prentice Hall, Englewood Cliffs, NJ, 2004).

\bibitem{merz} E. Merzbacher, {\it Quantum Mechanics} 
(John Wiley \& Sons, Hoboken, NJ, 1998).

\bibitem{sl1} E. Kreyszig, {\it Advanced Engineering Mathematics} (John Wiley \& Sons, 
Hoboken, NJ, 2006), pp. 188--193.

\bibitem{sl2} P. Dennery and A. Krzywicki, {\it Mathematics for 
Physicists} (Dover, Mineola, NY, 1995).

\bibitem{numerov} P. C. Chow, ``Computer solutions to the Schr\"{o}dinger 
equation,'' Am. J. Phys. {\bf 40}, 730--734 (1972).

\bibitem{landau} R. H. Landau, M. Jos\'{e} P\'{a}ez, and C. C. Bordeianu, 
{\it A Survey of Computational Physics} (Princeton University Press, 
Princeton, NJ, 2008).

\bibitem{ee} E. C. Titchmarsh, ``Some eigenfunction expansion formulae,'' 
Proc. London Math. Soc. {\bf 11}, 159--168 (1960) and references therein; 
C. E. Dean and S. A. Fulling, ``Continuum eigenfunction expansions and 
resonances: A simple model,'' Am. J. Phys. {\bf 50}, 540--544 (1982).

\bibitem{eemodern} See Chap. 4 in Refs.~\onlinecite{sl1,sl2}.

\bibitem{dirac} P. A. M. Dirac, {\it The Principles of Quantum Mechanics} 
(Oxford University Press, New York, 1996).

\bibitem{sakurai} J. J. Sakurai, {\it Modern Quantum Mechanics} (Addison -Wesley, Reading, MA, 1995).

\bibitem{scatter} A. Goldberg, H. M. Schey, and J. L. Schwartz, 
``One-dimensional scattering in configuration space and momentum space,'' 
Am. J. Phys. {\bf 36}, 454--455 (1968); 
S. K. Adhikari, ``Quantum scattering in two-dimensions,'' Am. J. Phys. 
{\bf 54}, 362--367 (1986).
\bibitem{onedelta} M. Lieber, ``Quantum mechanics in momentum space: An 
illustration,'' Am. J. Phys. {\bf 43}, 486--491 (1975).

\bibitem{landau2} L. D. Landau and E. M. Lifshitz, {\it Quantum 
Mechanics (Non-Relativistic Theory)} (Butterworth-Heinemann, Burlington, 
MA, 2005), Sec. 45.

\bibitem{twodelta1} J. Fernando Perez and F. A. B. Coutinho, ``Schr\"{o}dinger 
equation in two dimensions for a zero-range potential and a uniform magnetic 
field: An exactly solvable model,'' Am. J. Phys. {\bf 59}, 52--54 (1991).

\bibitem{twodelta2} S. Nyeo, ``Regularization methods for delta-function 
potential in two-dimensional quantum mechanics,'' Am. J. Phys. {\bf 68}, 
571--575 (2000).

\bibitem{coulomb} J. W. Huang and A. Kozycki, ``Hydrogen atom in two 
dimensions,'' Am. J. Phys. {\bf 47}, 1005--1006 (1979); X. L. Yang, M. Lieber, 
and F. T. Chan, ``The Runge-Lenz vector for the two-dimensional hydrogen 
atom,'' Am. J. Phys. {\bf 59}, 231--232 (1991); C. R. Jamell, C.-H. Zhang, and 
Y. N. Joglekar, ``Dilute excitons in a double-layer system: single exciton 
and mean-field approach,'' arXiv:0910.2993.

\bibitem{Ddim} L. E. Blumenson, ``A derivation of $n$-dimensional spherical 
coordinates,'' Am. Math. Monthly {\bf 67}, 63--66 (1960); E. Demiralp, 
``Bound states of $n$-dimensional harmonic oscillator decorated with Dirac 
delta functions,'' J. Phys. A {\bf 38}, 4783--4793 (2005). 

\end{thebibliography}
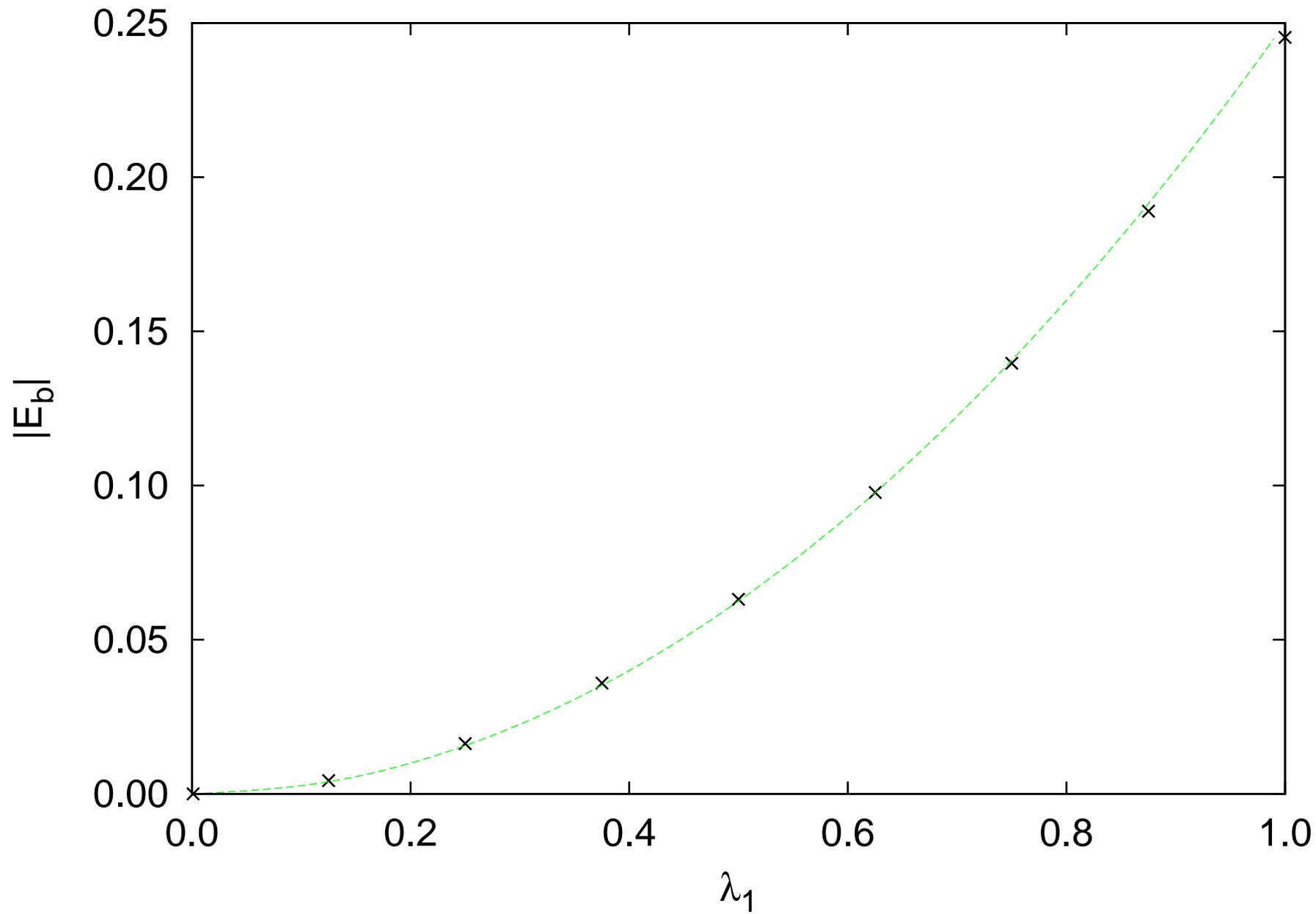
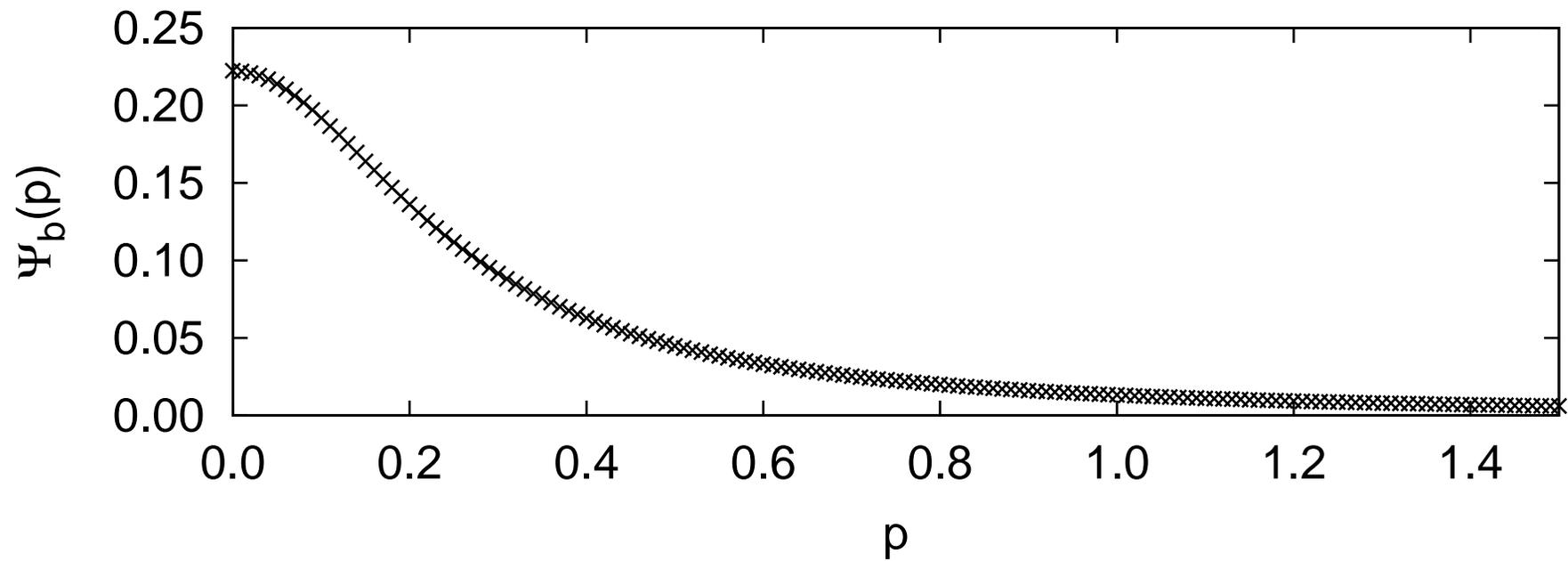
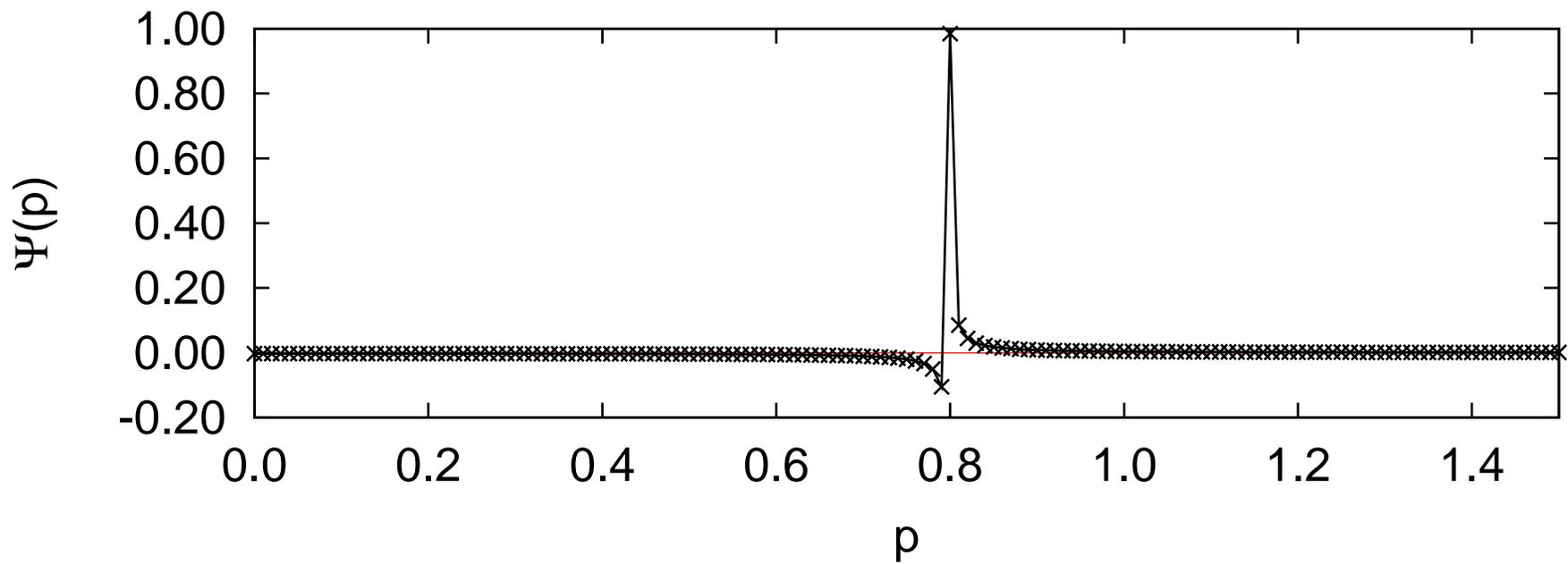
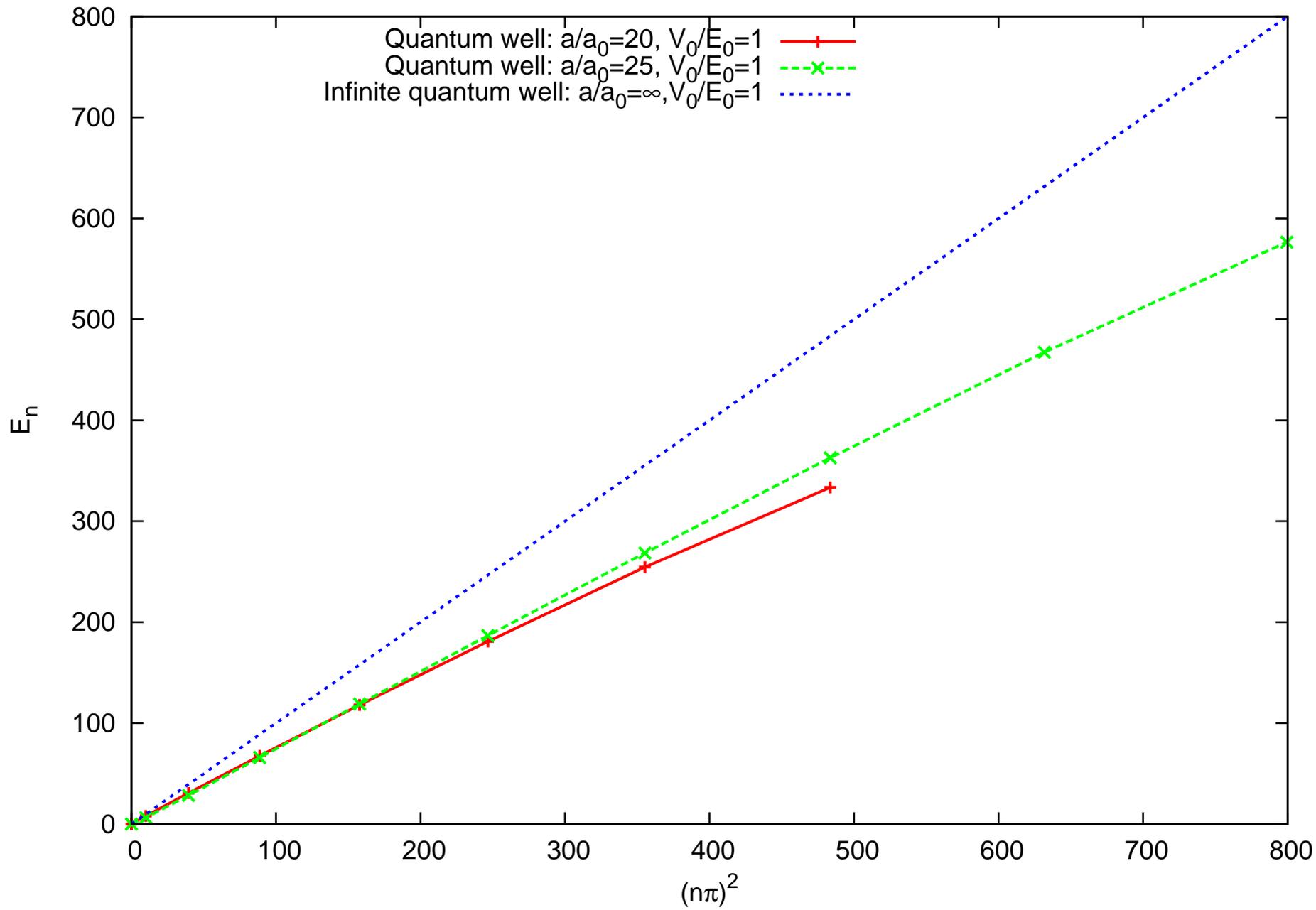
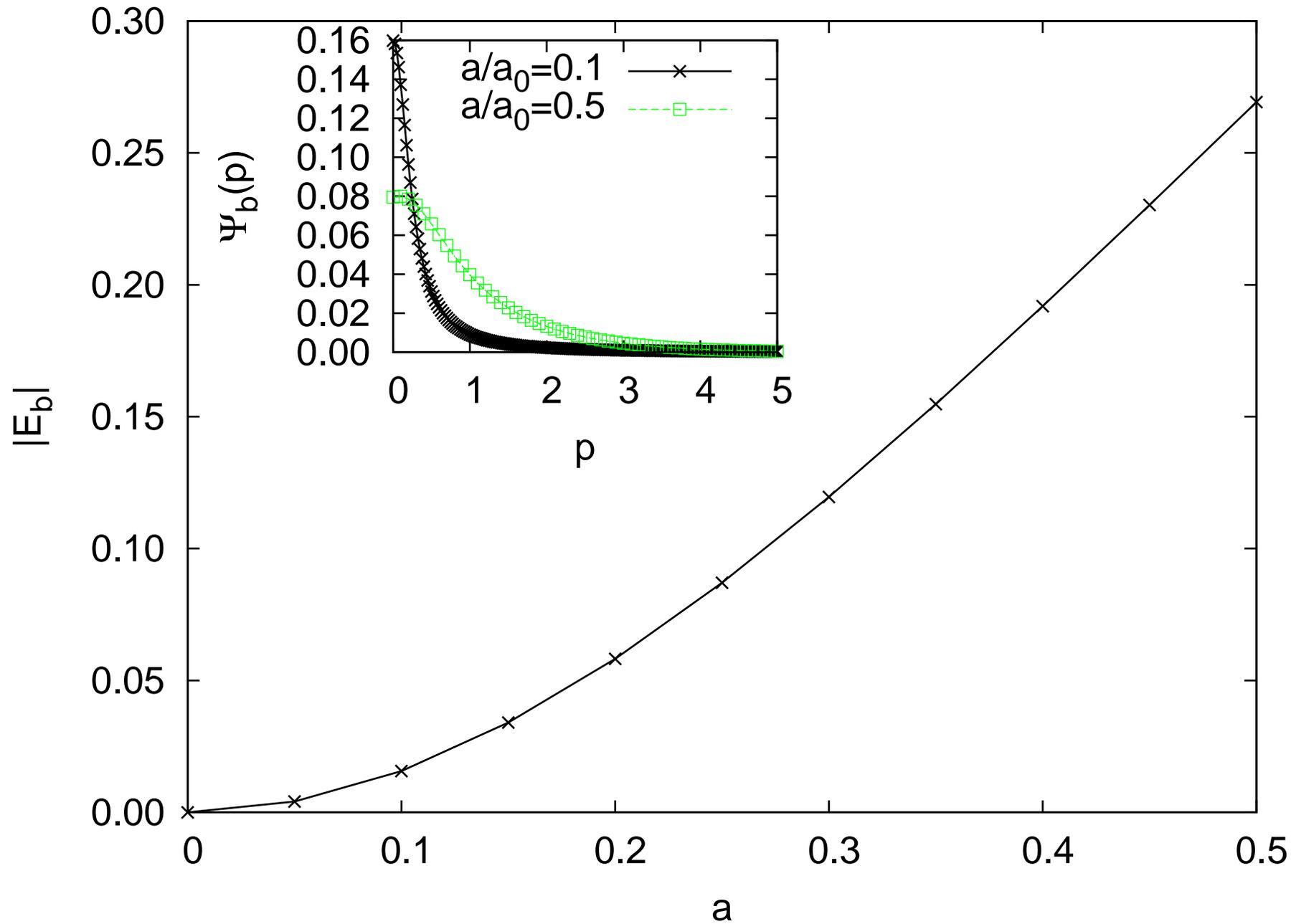
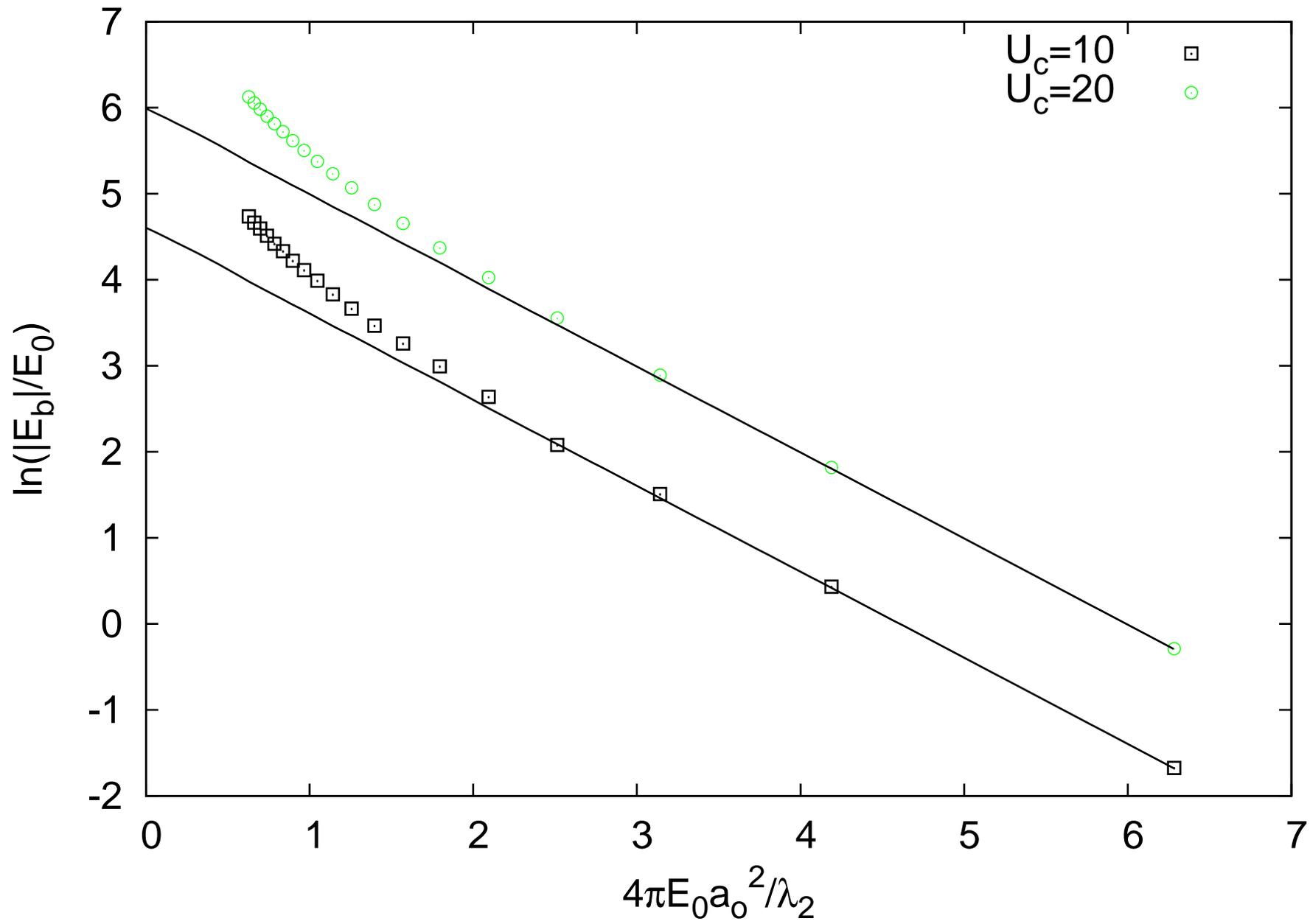
\end{document}